\documentclass[twocolumn,amsmath,amssymb,showpacs]{revtex4}
\usepackage{epsfig}
\usepackage{graphicx}

\begin{document}

\title{Possible quantum gravity effects on the gravitational deflection of light}

\author{Xin Li$^{1,3}$}
\email{lixin@itp.ac.cn}
\author{Zhe Chang$^{2,3}$}
\email{changz@ihep.ac.cn}
\affiliation{${}^1$Institute of Theoretical Physics,
Chinese Academy of Sciences, 100190 Beijing, China\\
${}^2$Institute of High Energy Physics, Chinese Academy
of Sciences, 100049 Beijing, China\\
${}^3$Theoretical Physics Center for Science Facilities, Chinese Academy of Sciences}

\begin{abstract}
We investigate possible quantum gravity (QG) effects on the gravitational deflection of light. Two forms of deformation of the Schwarzschild spacetime are proposed. The first ansatz is a given Finslerian line element, it could be regarded as a weak QG effect on the Schwarzschild spacetime. Starting from this ansatz, we deduce the deflection angle of the light ray which passes a weak gravitational source. The second ansatz could be regarded as a strong QG effect on the Schwarzschild spacetime. The deflection angle of the light ray which passes a weak gravitational source is deduced in this Riemannian spacetime. This QG effect may distinguish the mixed light rays in the absence of gravitational source by a ``spectroscope" (the gravitational source). The solutions of gravitational field equation in this Riemannian spacetime indicate that the QG effect could be regarded as the vacuum energy and the energy density of vacuum is related to the spacetime deformation parameter.
\end{abstract}

\maketitle
\section{Introduction}
The quantum theory of gravity (QG) has been studied for more than 70 years \cite{Stachel}. The most famous theories of quantum gravity, such as string theory \cite{string} and loop quantum gravity \cite{loop}, provide us some key features of QG.
Amelino-Camelia \cite{Amelino} listed possible candidates of QG effects: the violation of Lorentz symmetry of Standard Model \cite{Garay} as well as discrete symmetry (CPT symmetry) \cite{Ellis,Kostelecky}, Planck-scale fuzziness of spacetime \cite{Amelino1}. Other possible QG effects include: deviations from Newton¡¯s law at very short distances \cite{Hoyle}, possible production of mini-black holes \cite{Dimopoulos}.  The researches of such possible consequence of QG are refered as quantum gravity phenomenology (QGP). The rapid progress in technology makes experiments have the opportunities to test the sub-Planckian consequences of QG scenarios.

The QGP covers a wide range of subjects. One of the most important QG effects is the violation of the Lorentz invariance (LI)\cite{Mattingly}. A feature of QG, which is the most debated possibility for a quantum spacetime, manifests that the spacetime in Planck-scale may be noncommutative \cite{Connes,Majid}. A huge numbers of investigations of noncommutative spacetime manifest that the Lie-algebra Poincare symmetries are either broken to a smaller symmetry (Lie algebra or deformed into Hopf-algebra \cite{Majid1} symmetries). The LI violation implies that the dispersion relations for elementary particles should be modified. Moreover, studying on the dispersion relation is a convenient way for physicists to test the departure from LI. In the past few years,
Amelino-Camelia and Smolin as well as their collaborators have
developed the Doubly Special Relativity (DSR)
\cite{Amelino2,Amelino3,Amelino4,Smolin1,Smolin2} to take Planck-scale effects into
account by introducing an invariant Planckin parameter in the theory
of Special Relativity. The general form of dispersion relation for
free particles in the DSR is of the form
\begin{equation}
E^2=m^2+p^2+\sum_{n=1}^\infty\alpha_n(\mu,M_p)p^n~,
\end{equation}
where $p=\sqrt{\parallel \vec{p}\parallel^2}$, $\mu$ denotes a parameter of the theory with mass scale and
$M_p$ is the Planck mass. The modified dispersion relations (MDR)
have been tested through observations on gamma-ray bursts and
ultra-high energy cosmic rays\cite{Jacobson}. Girelli {\it et al.}\cite{Girelli} showed that the MDR can be incorporated into
the framework of Finsler geometry. The symmetry of the  MDR was
described in the Hamiltonian formalism. The generators of symmetry
commute with ${\cal M}(p)$ (here ${\cal M}(p)=m^2$ gives the mass
shell condition). The mass shell condition is invariant under the deformed
Lorentz transformations.

The research of Girelli {\it et al.} \cite{Girelli} gives a possible origin of MDR, which means the quantum spacetime may have a Finslerian form. Randers space, as a special kind of Finsler space, was first proposed by G. Randers \cite{Randers}. Within the framework of Randers space, modified dispersion relation has been discussed \cite{RF}, and the threshold of ultra high energy cosmic rays was investigated \cite{UHC}. The investigation of the isometric group of Finsler space indicates that the Lorentz symmetry is broken in Finsler space \cite{Li}.

Regardless of any particular theories of QG, we generally agree that the spacetime in QG scenario should be deformed. In this paper, we start from a giving deformed spacetime to investigate possible QG effects on the gravitational deflection of light. This paper is organized as follows. In Sec. II, we introduce a Finslerian line element which could be regarded as a weak deformation from the Schwarzschild spacetime. Starting from such line element, we investigate the trajectory of the light ray. In Sec. III, we introduce a Riemannian line element which could be regarded as a strong deformation from the Schwarzschild spacetime. The trajectory of the light ray in this Riemannian spacetime is studied. The Einstein's gravitational field equation is presented in this Riemannian spacetime. The light ray deflected by a weak gravitational source is studied. In Sec. IV, we give conclusions on the possible QG effects on the gravitational deflection of light.

\section{Weak quantum gravity effect on the gravitational deflection of light}
Instead of defining an inner product structure over the tangent bundle in Riemann geometry, Finsler geometry is based on
the so called Finsler structure $F$ with the property
$F(x,\lambda y)=\lambda F(x,y)$ for all $\lambda>0$, where $x$ represents position
and $y\equiv\frac{dx}{d\tau}$ represents velocity. The Finsler metric is given as\cite{Book
by Bao}
 \begin{equation}
 g_{\mu\nu}\equiv\frac{\partial}{\partial
y^\mu}\frac{\partial}{\partial y^\nu}\left(\frac{1}{2}F^2\right).
\end{equation}
Finsler geometry has its genesis in integrals of the form
\begin{equation}
\label{integral length}
\int^r_sF(x^1,\cdots,x^n;\frac{dx^1}{d\tau},\cdots,\frac{dx^n}{d\tau})d\tau~.
\end{equation}
So that the Finsler structure represents the length element of Finsler space.

Following the calculus of variations, one
get the geodesic equation of Finsler space\cite{Book by Bao}
\begin{equation}\label{geodesic of F}
\frac{d^2\sigma^\lambda}{d\tau^2}+\gamma^\lambda_{\mu\nu}\frac{d\sigma^\mu}{d\tau}\frac{d\sigma^\nu}{d\tau}=\frac{d\sigma^\mu}{d\tau}\frac{d}{d\tau}\left(\log
F(\sigma, \frac{d\sigma}{d\tau})\right),
\end{equation}
where
\begin{equation}
\gamma^\lambda_{\mu\nu}=\frac{g^{\lambda\alpha}}{2}\left(\frac{\partial g_{\mu\alpha}}{\partial x^\nu}+\frac{\partial g_{\nu\alpha}}{\partial x^\mu}-\frac{\partial g_{\mu\nu}}{\partial x^\alpha}\right)
\end{equation}
is the formal Christoffel symbols of the second kind with the same form of Riemannian connection.
The parallel transport
has been studied in the framework of Cartan
connection\cite{Matsumoto,Antonelli,Szabo}. The notation of parallel
transport in Finsler manifold means that the length
$F\left(\frac{d\sigma}{d\tau}\right)$ is constant. Thus, the
autoparallel equation can be got from the equation (\ref{geodesic of
F})
\begin{equation}\label{autoparallel}
\frac{d^2\sigma^\lambda}{d\tau^2}+\gamma^\lambda_{\mu\nu}\frac{d\sigma^\mu}{d\tau}\frac{d\sigma^\nu}{d\tau}=0.
\end{equation} Since the geodesic equation (\ref{geodesic of F}) is directly
derived from the integral length of $\sigma$
\begin{equation} L=\int
F\left(\frac{d\sigma}{d\tau}\right)d\tau,
\end{equation} the inner product
$\left(\sqrt{g_{\mu\nu}\frac{d\sigma^\mu}{d\tau}\frac{d\sigma^\nu}{d\tau}}=F\left(\frac{d\sigma}{d\tau}\right)\right)$
of two parallel transported vectors is preserved.

In a phenomenological approach, it is reasonable to suppose that the line element of spacetime is given as
\begin{eqnarray}
\label{ansatz1}
F^2d\tau^2=&&adt^2-a^{-1}dr^2-r^2(\sin^2\theta d\theta^2+d\phi^2)\nonumber\\
&+&\kappa \sqrt{a}\frac{GM}{r}\left(1-a\frac{J^2}{E^2 r^2}\right)^{\frac{3}{4}}dt\sqrt{dtdr},
\end{eqnarray}
where $a=1-\frac{2GM}{r}$, $\frac{GM}{r}\ll1$, $M$ is the mass of gravitational source, $E$ is the energy per unit mass of the particle, $J$ is the angular momentum per unit mass of the particle and $\kappa$ is dimensionless and constancy parameter which is function of the new physics scale.
The dimensionless parameter $\kappa$ just plays the role of the measurement of QG effect. While $\kappa$ vanishes, the line element (\ref{ansatz1}) returns to the famous Schwarzschild metric. The line element (\ref{ansatz1}) could be regarded as the deformation of the Schwarzschild metric. Since the coefficient of the non Riemannian term in (\ref{ansatz1}) is isotropic, it is convenient to consider that the motion of particle is confined in the plane of $\theta=\frac{\pi}{2}$. Therefore, the trajectory of the particle is described by the $r$ and $\phi$ coordinates, and the line element reduces to
\begin{eqnarray}\label{ansatz2}
F^2d\tau^2=&&adt^2-a^{-1}dr^2-r^2d\phi^2\nonumber\\
&+&\kappa \sqrt{a}\frac{GM}{r}\left(1-a\frac{J^2}{E^2 r^2}\right)^{\frac{3}{4}}dt\sqrt{dtdr}.
\end{eqnarray}

It should be noticed that the coefficients of the line element (\ref{ansatz2}) do not depend either on $r$ or $\phi$. By making use of the autoparallel geodesic equation (\ref{autoparallel}), we obtain two integrals of motion
\begin{eqnarray}\label{motion1}
a\dot{t}+\kappa \sqrt{a}\frac{3GM}{4r}\left(1-a\frac{J^2}{E^2 r^2}\right)^{\frac{3}{4}}\sqrt{\dot{t}\dot{r}}&=&E,\\
\label{motion2}
r^2\dot{\phi}&=&J,
\end{eqnarray}
where a dot denotes the derivatives with respect to trajectory parameter $\tau$. The deformation term in (\ref{ansatz2}) is very small, here we can approximately choose the two integral constants to be $E$ and $J$, respectively. In this paper, we mainly consider the motion of light ray. There is an additional constraint arising from $F=0$ for the motion of light ray
\begin{equation}\label{motion3}
a\dot{t}^2-a^{-1}\dot{r}^2-r^2\dot{\phi}^2+\kappa \sqrt{a}\frac{GM}{r}\left(1-a\frac{J^2}{E^2 r^2}\right)^{\frac{3}{4}}\dot{t}^{\frac{3}{2}}\dot{r}^{\frac{1}{2}}=0.
\end{equation}
By making use of the equations (\ref{motion1}), (\ref{motion2}) and (\ref{motion3}), we obtain an approximate equation for $\dot{r}$, which is valid for $\frac{GM}{r}\ll1$ (it was assumed in the beginning of this section)
\begin{equation}\label{eq dotr}
\dot{r}=\sqrt{E^2-a\frac{J^2}{r^2}}\left(1-\kappa\frac{GM}{4r}\right).
\end{equation}
Combining (\ref{eq dotr}) with equation (\ref{motion2}), we obtain
\begin{equation}\label{eq dotr1}
\left(\frac{1}{r^2}\frac{dr}{d\phi}\right)^2=\left(\frac{E^2}{J^2}-\frac{a}{r^2}\right)\left(1-\kappa\frac{GM}{2r}\right).
\end{equation}
In terms of the variable $u=\frac{GM}{r}$, the equation (\ref{eq dotr1}) changes as
\begin{equation}\label{eq u}
\left(\frac{du}{d\phi}\right)^2=\left(\left(\frac{EGM}{J}\right)^2-u^2(1-2u)\right)\left(1-\frac{\kappa u}{2}\right).
\end{equation}
The derivatives $\frac{d}{d\phi}$ of the equation (\ref{eq u}) gives
\begin{equation}\label{eq u1}
\frac{d^2u}{d\phi}^2+u=3u^2-\frac{\kappa}{4}\left(\left(\frac{EGM}{J}\right)^2-3u^2\right)+O(u^3).
\end{equation}
Solving the equation (\ref{eq u1}) to order $u$, we get that
\begin{equation}\label{FO solu}
u=u_0\cos\phi.
\end{equation}
Substituting the above solution into (\ref{eq u}), we obtain the approximate solution for $u_0$
\begin{equation}\label{u0}
u_0=\frac{EGM}{J}=\frac{GM}{\xi},
\end{equation}
where $\xi$ is the minimum distance of the light ray to the gravitational source with mass $M$.
The closest approach of the light ray to the gravitational source implies $\frac{dr}{d\phi}=0$. Then, deducing from the equation (\ref{eq dotr1}), we obtain approximate solution for $\xi$$(=\frac{J}{E}$). This is the reason for the second equation in (\ref{u0}).
By making use of the first order approximation (\ref{FO solu}), one may get the second order approximation of (\ref{eq u1})
\begin{equation}
u=u_0^2\left((1+\kappa/4)(1+\sin^2\phi)-\kappa/4\right).
\end{equation}
Thus, the solution of equation (\ref{eq u1}) is
\begin{equation}\label{solu u}
u=u_0\cos\phi+u_0^2\left((1+\kappa/4)(1+\sin^2\phi)-\kappa/4\right).
\end{equation}
At infinity ($u=0$), the solution (\ref{solu u}) shows that the angle $\phi=\pm(\frac{\pi}{2}+\alpha)$, and the small angle $\alpha$ satisfies the constraint
\begin{equation}\label{alpha}
-u_0\sin\alpha+u_0^2\left((1+\kappa/4)(1+\cos^2\alpha)-\kappa/4\right)=0.
\end{equation}
Since the angle $\alpha$ is very small, the solution of the above equation (\ref{alpha}) is
\begin{equation}
\alpha=u_0(2+\kappa/4)=(2+\kappa/4)\frac{GM}{\xi}.
\end{equation}
The two asymptotic directions differs from $\pi$ by the deflection angle
\begin{equation}\label{df angle}
\hat{\alpha}=2\alpha=(4+\kappa/2)\frac{GM}{\xi}
\end{equation}

In general relativity, the light passing a massive object $M$ at a minimum distance $\xi$ suffers deflection, and the deflection angle (``Einstein angle") is $\hat{\alpha}_E=\frac{4GM}{\xi}$ \cite{Weinberg}. The spacetime (\ref{ansatz1}) deformed from Schwarzschild spacetime implies a deformed deflection angle. The formula for the deformed deflection angle (\ref{df angle}) differs from $\hat{\alpha}_E$ by $\kappa\frac{GM}{2\xi}$, which is proportional to the deformed parameter $\kappa$.

In astronomical observations, the gravitational lensing surveys is used to calculate the mass distribution that projected
onto the sky. A large amount of observations manifest that the expected gravitational lensing effects deducing by the Einstein angle are not in accord with the experimental data. An example is the full-sky data product for the Bullet Cluster 1E0657-558 \cite{Clowe}.   We wish the deformed deflection angle (\ref{df angle}) may account to these observations.

\section{Strong quantum gravity effect on the gravitational deflection of light}
In Sec. II, we got a deformed deflection angle for the light ray. The Finslerian metric (\ref{ansatz1}) could be regarded as a weak deformation from the Schwarzschild metric. In this section, we discuss the strong quantum gravity effect on the gravitational deflection of light. Again, in a phenomenological approach, we propose that the line element of spacetime is given as
\begin{equation}\label{ansatz3}
ds^2=adt^2-(4n+1)^2a^{-1}dr^2-r^2d\phi^2,
\end{equation}
where the deformation parameter $n=0,1,2\cdots$. In ansatz (\ref{ansatz3}), we already confined the motion of particle in the plane of $\theta=\frac{\pi}{2}$. The reason is the same within Sec. II. The ansatz (\ref{ansatz3}) manifests a large deformation from the Schwarzschild metric, it returns to the Schwarzschild metric while the deformation parameter vanishes.

%Since the coefficients of the line element (\ref{ansatz3}) do not depend neither on $r$ or $\phi$,
By making use of the autoparallel geodesic equation of Riemannian space \cite{Weinberg}, we obtain two integrals of motion
\begin{eqnarray}\label{motion st1}
a\dot{t}&=&E,\\
\label{motion st2}
r^2\dot{\phi}&=&J.
\end{eqnarray}
The additional constraint arising from $\frac{ds}{d\tau}=0$ for the motion of light is
\begin{equation}\label{motion st3}
a\dot{t}^2-(4n+1)^2a^{-1}\dot{r}^2-r^2\dot{\phi}^2=0
\end{equation}
By making use of the equations (\ref{motion st1}), (\ref{motion st2}) and (\ref{motion st3}), we get an approximate equation for $\dot{r}$
\begin{equation}\label{st dotr}
(4n+1)^2\dot{r}^2=E^2-a\frac{J^2}{r^2}.
\end{equation}
Combining this equation (\ref{st dotr}) with (\ref{motion st2}), we obtain that
\begin{equation}\label{st dotr1}
\left(\frac{4n+1}{r^2}\frac{dr}{d\phi}\right)^2=\frac{E^2}{J^2}-\frac{a}{r^2}.
\end{equation}
The closest approach of the light ray to the gravitational source $M$ implies $\frac{dr}{d\phi}=0$. Then, deducing from the equation (\ref{st dotr1}), we obtain approximate solution for the distance of closest approach $\xi$($=\frac{J}{E}$).
Changing variable to $u=\frac{GM}{r}$, we obtain from equation (\ref{eq dotr1}) that
\begin{equation}\label{steq u}
(4n+1)^2\left(\frac{du}{d\phi}\right)^2=\left(\frac{EGM}{J}\right)^2-u^2(1-2u).
\end{equation}
Calculating the derivatives $\frac{d}{d\phi}$ of the equation (\ref{steq u}) gives
\begin{equation}\label{steq u1}
\frac{d^2u}{d\phi^2}+\frac{u}{(4n+1)^2}=\frac{3u^2}{(4n+1)^2}.
\end{equation}
Noticing that $u=\frac{GM}{r}\ll1$, the equation (\ref{steq u1}) has solution
\begin{equation}\label{stsolu u}
u=u_0\cos\frac{\phi}{4n+1}+u_0^2\left(1+\sin^2\frac{\phi}{4n+1}\right).
\end{equation}
Substituting the above solution into (\ref{steq u}), we get the approximate solution for $u_0$
\begin{equation}\label{st u0}
u_0=\frac{EGM}{J}=\frac{GM}{\xi}.
\end{equation}
At infinity ($u=0$), to first order in $u_0$, the solution (\ref{stsolu u}) shows
\begin{equation}\label{phi0}
\phi=\pm(4n+1)\frac{\pi}{2}.
\end{equation}
The angle $\phi$ in (\ref{phi0}) for different $n$ only differ from an integer times of $2\pi$. The formula (\ref{phi0}) means each light ray which corresponds to a given parameter $n$ is mixed at infinity in the absence of the gravitational source $M$, and the light rays all move with the same direction. When they suffer from a week gravitational source, to second order in $u_0$, the solution (\ref{stsolu u}) implies
\begin{equation}\label{phi1}
\phi=\pm\left((4n+1)\frac{\pi}{2}+\alpha\right),
\end{equation}
and the small angle $\alpha$ satisfies the constraint
\begin{equation}\label{alpha st}
-u_0\sin\frac{\alpha}{4n+1}+u_0^2\left(1+\cos^2\frac{\alpha}{4n+1}\right)=0.
\end{equation}
Since the angle $\alpha$ is very small, the solution of the above equation (\ref{alpha st}) is
\begin{equation}
\alpha=2(4n+1)u_0=2(4n+1)\frac{GM}{\xi}.
\end{equation}
The difference of two asymptotic directions $\pm\left((4n+1)\frac{\pi}{2}+\alpha\right)$ differs from $(4n+1)\pi$ by the deflection angle
\begin{equation}\label{df angle st}
\hat{\alpha}=2\alpha=4(4n+1)\frac{GM}{\xi}.
\end{equation}
This formula (\ref{df angle st}) means the deflection angles for different $n$ are different. Therefore, the weak gravitational source $M$ just plays the role of the ``spectroscope". The mixed light rays come from infinity, refracted by the ``spectroscope" $M$ at the same distance of closest approach $\xi$, go into different directions. And the deflection angle is in proportion to the spacetime deformation parameter $n$.

One should notice from the formula (\ref{st dotr}) that the radial momentum of light rays is modified. The ansatz (\ref{ansatz3}) is a Riemannian line element. Therefore, the motion of particle in this spacetime satisfies the Einstein's gravitational field equation. The Einstein's gravitational field equation is taken the from
\begin{equation}\label{field eq}
R_{\mu\nu}=-8\pi G\left(T_{\mu\nu}-\frac{1}{2}g_{\mu\nu}T^\lambda_{~\lambda}\right),
\end{equation}
where $R_{\mu\nu}$ is the Ricci tensor.
The energy-momentum tensor of a perfect fluid is taken the from as
\begin{equation}
T^{\mu\nu}=-pg^{\mu\nu}+(p+\rho)U^\mu U^\nu,
\end{equation}
here $U^\mu$ is the fluid four-velocity and satisfies $g_{\mu\nu}U^\mu U^\nu=1$ , $p$ and $\rho$ are pressure and energy density respectively.
By making use of the line element (\ref{ansatz3}), we show that the combination of the $tt$ and $rr$ component of the field equation (\ref{field eq}) gives
\begin{equation}\label{field solu eq1}
p=-\rho,
\end{equation}
and the $\phi\phi$ component of the field equation (\ref{field eq}) gives
\begin{equation}\label{field solu eq2}
-1+\frac{1}{(4n+1)^2}=-4\pi Gr^2(\rho-p).
\end{equation}
The equation (\ref{field solu eq1}) implies that there is a quantum vacuum with energy density $\rho$ outside the gravitational source. Combining the equation (\ref{field solu eq1}) with (\ref{field solu eq2}), we obtain
\begin{equation}\label{field solu eq3}
\rho=\frac{1}{8\pi Gr^2}\left(1-\frac{1}{(4n+1)^2}\right).
\end{equation}
The equation (\ref{field solu eq3}) indicates that the energy density of vacuum is the function of spacetime deformation parameter $n$. From the point of view of particle physics, the cosmological constant, a popular candidate of the dark energy, naturally arises as an energy density of the vacuum \cite{Copeland}. It implies that this kind of QG effect (\ref{ansatz2}) may arise from the influence of the dark energy.

\section{Conclusion}
In this paper, we investigated the possible quantum gravity effects on the gravitational deflection of light. One of the most expected QG effect is the deformation of spacetime geometry. In a phenomenological approach, we proposed two deformations of the Schwarzschild spacetime. The first ansatz (\ref{ansatz1}) is a given Finslerian line element, it could be regarded as a weak QG effect on the Schwarzschild spacetime. The deformation term (the non Riemannian term) is very small. Starting from the ansatz (\ref{ansatz1}), we deduced the deflection angle (\ref{df angle}). This deformed deflection angle may account to the observations of gravitational lensing which can not be explained in the framework of general relativity.

The second ansatz (\ref{ansatz3}) could be regarded as a strong QG effect on the Schwarzschild spacetime, for the deformation term $(4n+1)^2\geq1$. Starting from the ansatz (\ref{ansatz3}), we deduced the deflection angle (\ref{df angle st}). This QG effect may distinguish the mixed light rays in the absence of gravitational source by a ``spectroscope" $M$. The solutions of gravitational field equation in spacetime (\ref{ansatz3}) indicate that the QG effect could be regarded as the vacuum effect and the energy density of vacuum is related to the spacetime deformation parameter $n$. While $n$ vanishes, the spacetime with quantum effect (\ref{ansatz3}) returns to the Schwarzschild spacetime and the vacuum energy vanishes, it means that there is nothing exist outside the gravitational source $M$.

\bigskip

\centerline{\large\bf Acknowledgements} \vspace{0.5cm}
 We would like to thank Prof. C. J. Zhu for useful discussions. The
work was supported by the NSF of China under Grant No. 10525522 and
10875129.

\end{document}